\documentclass[lettersize,journal]{IEEEtran}
\usepackage{amsmath,amsfonts,amssymb}
\usepackage{algorithmic}
\usepackage{algorithm}
\usepackage{array}
\usepackage[caption=false,font=normalsize,labelfont=sf,textfont=sf]{subfig}
\usepackage{textcomp}
\usepackage{stfloats}
\usepackage{url}
\usepackage{verbatim}
\usepackage{graphicx}
\usepackage{cite,xcolor}
\usepackage{multirow,multicol}
\usepackage{balance}
\usepackage{cleveref}
\begin{document}

\title{\Huge{Communicating in the Mediumband:}\\What it is and Why it Matters}

\author{Dushyantha A Basnayaka,~\IEEEmembership{Senior Member,~IEEE}
%
%
}
%
\markboth{IEEE Communications Magazine, Accepted, Jul. 2024.}%
{Shell \MakeLowercase{\textit{et al.}}: A Sample Article Using IEEEtran.cls for IEEE Journals}
\begin{figure*}[t]
	\centerline{\includegraphics*[scale=1]{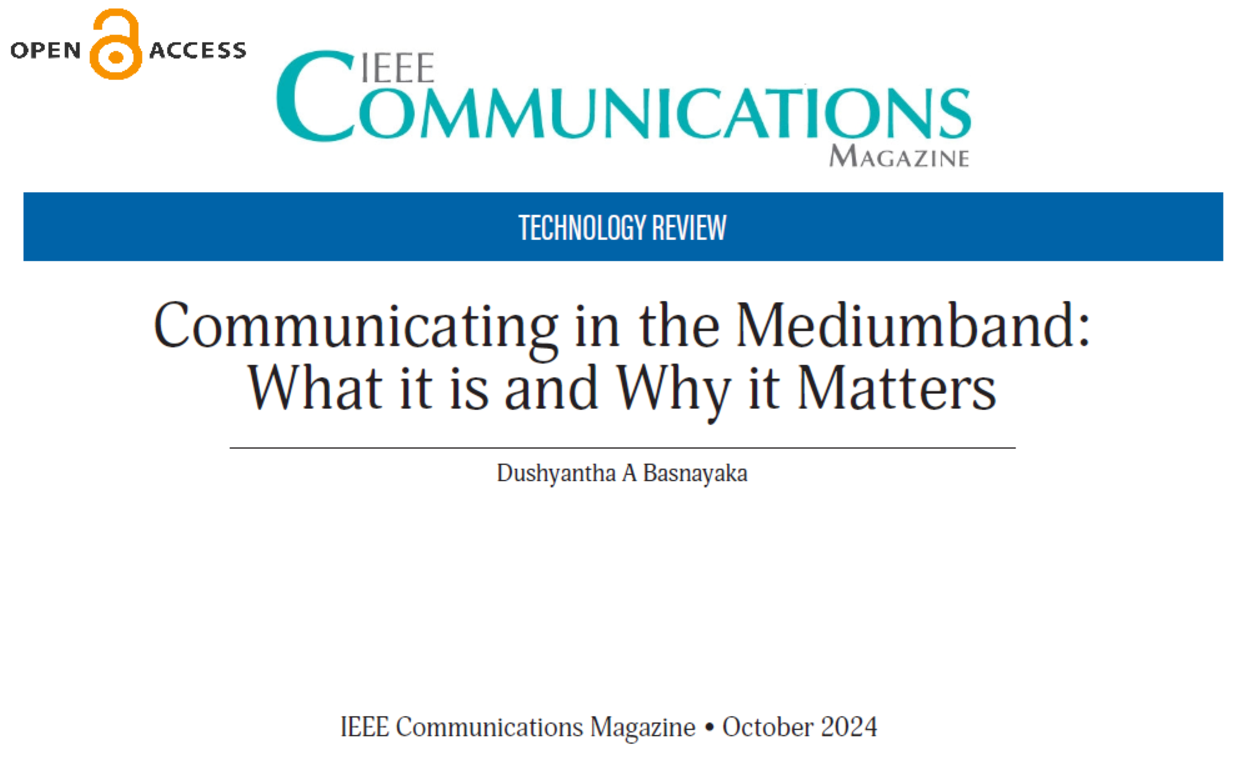}}
	\hrule
\vspace{5mm}
\textcolor{red}{\textit{Note: This article has been accepted for inclusion in a future issue of this magazine, and this is the author's version which has not been fully edited and content may change prior to final publication.}}
\end{figure*}
\maketitle
\vspace{-5mm}
\begin{abstract}
This paper, based on recent research, articulates the opportunities and challenges posed by an emerging area of study known as ``\textit{mediumband wireless communication}'', which refers to digital radio-frequency (RF) wireless communication through mediumband channels. This class of channels that falls in the transitional region between the narrowband and broadband channels, in many ways, is unique and shows significant potential. For instance, the effect of a highly unfavourable non-line-of-sight (NLoS) propagation environment can be transformed into a significantly favourable condition without making any intervention on the original propagation environment, but by simply communicating in the mediumband. The more unfavourable a propagation environment for wireless communication, the higher the potential gain by communicating in the mediumband. In this paper, using lay language as much as possible, we elaborate the unique properties of mediumband channels and implications of communicating in the mediumband for wider wireless communication along with some future research directions.
\end{abstract}


\section{Introduction}
\vspace{0mm}
Modern wireless communication has been serving us very well. You, me, and many billions of people around the world in every single second are benefited from it. For instance, when one taps ``\textit{send}'' button on a chat app on their smartphone, when a group of friends making a conference call from their mobile phones, when one watches their favourite movie on-the-go, or when one makes a call travelling at high speed, it is the brilliance of modern digital wireless communication built into thier mobile phones that they experience. For the continual economic growth however, the current capabilities of the global wireless communication is wholly inadequate. Hence, to create and inspire innovative business models, the enabling technologies like wireless communication should be made to be as ubiquitous as possible, as reliable as possible, as fast as possible, and also as efficient as possible.\\
\indent Strangely, wireless communication is, not only a business between the wireless transmitter (TX) and the wireless receiver (RX), but also inextricably intertwined with the propagation environment. Thanks to many pioneers before us, today’s wireless TXs and RXs are so advanced, and operate at the very edge of human ingenuity. In today's wireless communication, the effect of the propagation environment on wireless communication is the main limiting factor. The wireless communication would be reliable and can be made to be more efficient in terms of energy and spectral usage if the propagation environment, which is the entity between the TX and RX, is favourable for wireless communication (see \cref{xxx}). If it is unfavourable, on the other hand, the communication would be affected adversely. If the propagation environment is extremely unfavourable, which is not improbable, the communication would completely break down even with the state-of-the-art TXs and RXs.\\
\begin{figure}[t]
	\centerline{\includegraphics*[scale=0.75]{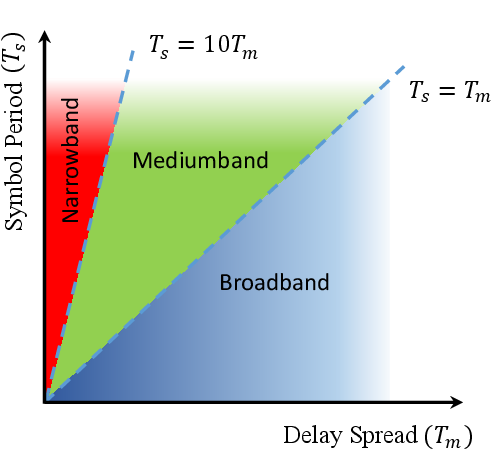}}
	\caption{Main regions on the $T_mT_s$--plane representing three classes of wireless channels. \cite{BasJ23}.}\label{fig:fig0}
	\vspace{-5mm}
\end{figure}
\indent Against such a backdrop, communicating in the mediumband, or otherwise mediumband wireless communication, has been shown to offer us refreshingly new perspectives to mitigate the adverse effects of the propagation environment enabling future wireless communication systems to achieve unprecedented gains in terms of reliability and efficiency \cite{BasJ23,Bas22}. Unlike technologies like intelligent reflective surfaces (IRS) \cite{ZZhang23}, mediumband wireless communication achieves this not by making any physical interventions, or introducing any foreign objects in, to the propagation environment. Importantly, it is the effect of the propagation environment, but not the propagation environment itself, that this new technology alters. What does this ``\textit{\textbf{mediumband}}'' really mean?\\
\indent The ``\textit{mediumband}'' is a new term that itself does not mean much. It does \textit{NOT} refer to any particular band of frequencies in the electromagnetic spectrum like microwave band, mmwave band or terahertz band. Collectively though, mediumband wireless communication refers to digital wireless communication through something called mediumband channels\footnote{Similarly, narrowband and broadband wireless communication refer to wireless communication respectively through narrowband and broadband channels. Importantly, it is the type of the channel that defines the type of the communication. Note that the term ``\textit{\textbf{channel}}'' referred herein is an abstract concept represented by a point on the $T_mT_s$-plane in Fig. \ref{fig:fig0}, where mediumband channels are represented by the points in the GREEN region. The carrier frequency has nothing to do with the definition of wireless channels.}, which is a newly identified class of channels that falls in the transitional region between the widely known classes of narrowband and broadband channels as shown in Fig. \ref{fig:fig0} \cite{BasJ23}. This class of mediumband channels has been shown to have unique properties showing significant potential to enhance the capabilities of future wireless communication on multiple fronts \cite[Sec. I-C]{BasJ23}.\\ 
\begin{figure*}[t]
	\centerline{\includegraphics[scale=0.8]{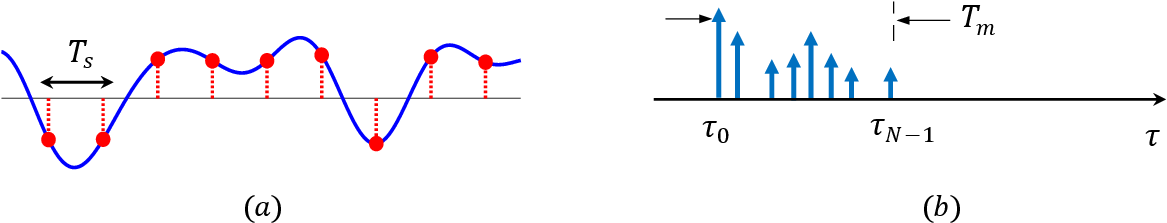}}
	\caption{a) A depiction of a typical transmit signal $s(t)$. In digital wireless communication, not all parts of $s(t)$, but only the points regularly separated in time (i.e. RED dots), carry information. This separation time is the symbol period or $T_s$. b) A depiction of a multipath profile as an impulse train, where the $x$--coordinate and the height of the impulses represent the excess delay and the strength of the corresponding MPC respectively. $T_m$ is a suitable measure of the time spread of MPCs. A wireless communication system, whose $T_m$ and $T_s$ approximately satisfy the condition: $0.1T_s \leq T_m \leq T_s$, which is represented by the GREEN region in Fig. \ref{fig:fig0}, is said to be operating in the mediumband.}\label{fig:fig21}
	\vspace{-5mm}
\end{figure*}
\indent This article, based on the recent research, highlighting the unique properties of mediumband channels, reviews the opportunities and challenges posed by mediumband wireless communication as an enabling communication technology for future wireless communication.
In particular, the key messages of this review are that:
\begin{itemize}
\item Mediumband wireless communication, among other benefits, enables us to alter the effect of the wireless propagation environment on wireless communication. For instance, the effect of a highly unfavourable non-line-of-sight (NLoS) propagation can be transformed into a significantly favourable condition for wireless communication without making any physical intervention to the original propagation environment.
\item  Wireless systems operating in the mediumband experiences increased level of inter-symbol-interference (ISI) reducing the channel quality in terms of signal-to-interference-ratio (SIR) or signal-to-interference-plus-noise-ratio (SINR). However, on the contrary to the widely held belief, it is not entirely the average SIR (or SINR) that dictates the performance of wireless communication in the mediumband, but the net effect of the SIR and another favourable effect, that could successfully counter the adverse effect of the increased level of ISI, known as the “\textit{\textbf{effect of deep fading avoidance}}”
\item  As a rule of thumb, the more unfavourable a propagation environment for wireless communication,
the higher the potential gain by communicating in
the mediumband.
\item The probability density function (PDF) of the desired fading factor in the mediumband channels in NLoS propagation has a unique bimodal distribution (see Figs \ref{fig:fig2}, \ref{fig:fig31}, \ref{fig:fig41}, and \ref{fig:fig4}). This PDF can be accurately modelled as an analytically amenable Gaussian difference rule (i.e. (6)), which seems to be able to predict the link-level performance of mediumband channels quite accurately.
\item The mediumband region (i.e., in GREEN) on the $T_mT_s$-plane in Fig. \ref{fig:fig0} is distinct and has unique properties like the ``\textit{\textbf{four leaf clover shape}}'' of the scatter plot, and the bimodal nature of the PDF, of the desired fading factor in the mediumband channels. So the conventional teaching of wireless communication may need revisions to include, not only the narrowband and broadband classes, but also the class of mediumband highlighting its fundamental differences and their wider implications for wireless communication.
\item The impact of communicating in the mediumband for wider wireless communication may be deep and wide. When the constituent links in systems like multiple antenna (i.e., MIMO), cognitive radio \cite{Arslan09}, relaying\cite{Pabst04}, reflective intelligent surfaces (RISs) \cite{ZZhang23}, space-time coding, limited feedback, non-orthogonal multiple access (NOMA) \cite{Ding17}, rate-splitting multiple access \cite{Mao22}, integrated sensing and communication (ISAC) \cite{Zhang21}, and countless emerging application areas of RF wireless communication are made to operate in the mediumband, these systems are expected to offer significant performance gains and create new opportunities. The extent of the potential gains and the challenges are, however, not yet fully known. The tools and the insights presented in this paper and its key references may be useful and a good starting point.
\end{itemize} 
\indent The rest of the paper is organized as follows. Starting with the preliminaries required for our discussion in Sec. \ref{sec:prelim}, Sec. \ref{sec:deep_fading} discusses the single most problematic phenomenon in wireless communication that is deep fading. The key elements of mediumband channels are elaborated in Sec. \ref{sec:mediumband} followed by a discussion on the effect of deep fading avoidance in Sec. \ref{sec:dfa}. Sec. \ref{sec:numerical_results} presents some emblematic numerical results based on a detailed link-level simulation, and Sec. \ref{sec:statistical_model} outlines a new unified statistical model to model the fading process in mediumband channels. The wider impact and challenges posed by mediumband wireless communication are discussed in Sec. \ref{sec:impact_problems} followed by a conclusion in Sec. \ref{sec:conclusion}.
\begin{figure*}[t]
	\centerline{\includegraphics*[scale=0.85]{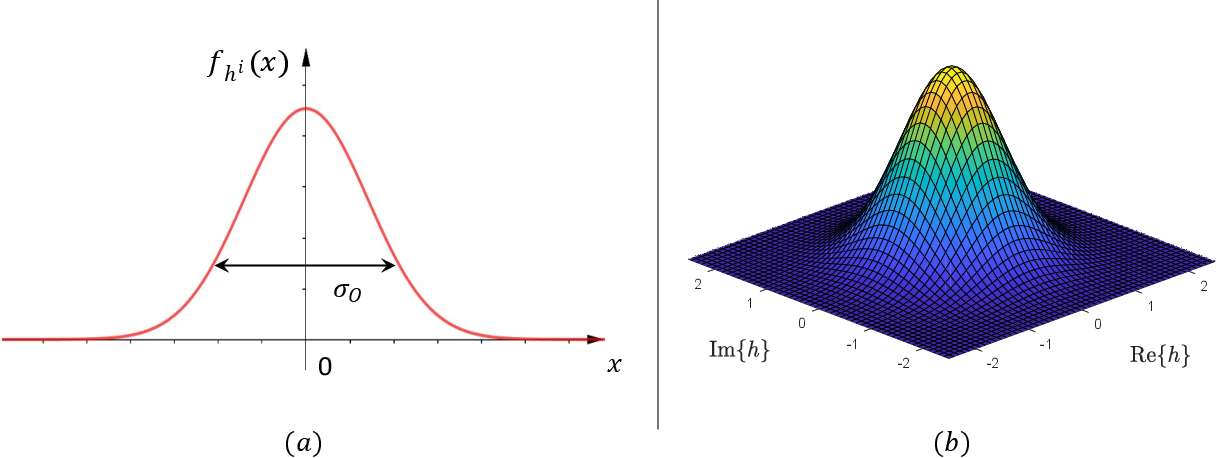}}
	\caption{a) A graphical depiction of the PDF of $h$ in a typical NLoS propagation environment as described in (3). Typically, $h$ is complex, but here only the PDF of the real part of $h$ denoted by $h^i$, that is $f_{h^i}(x)$, is shown for brevity. Note that $\sigma_O$ is a parameter to capture the width of the PDF, but does not measure the distance as shown. b) A depiction of $f_{h}(x,y)=f_{h^i}(x)f_{h^q}(y)$, which is the complex PDF of $h$. Note the peak at (0,0) signifying the fact that the probability of either Re$\{h\}$ or Im$\{h\}$ being very small is very high, which is the single most problematic dampening force in modern wireless communication.}\label{fig:fig1}
\end{figure*}
\section{Preliminaries}\label{sec:prelim}
In baseband equivalent form, the receive signal $r(t)$ at a typical wireless RX, due to the effect of multipath, is a random mixture of delayed and attenuated versions of the information bearing transmit signal $s(t)$, which can be mathematically given by:
\begin{align} \label{eq11}
	r(t) &= \underbrace{\sum _{n=0}^{N-1} \alpha _{n} e^{-j\phi _{n}} s\left ({t-\tau _{n}}\right )}_{r'(t)} + w(t),
\end{align}
where $\alpha _{n}$, $\phi _{n}$ and $\tau _{n}$ respectively known as the path gain, path phase and path delay of the $n$th multipath component (MPC) \cite[eq. 1]{Turin72}. Here, $w(t)$ is the ever-present noise in additive form. The actual values of $\alpha _{n}$, $\phi _{n}$ and $\tau _{n}$ are dependent on the propagation environment between the TX and the RX. As shown in Fig. \ref{fig:fig21}, in $s(t)$, only the signal points regularly separated in time carry digital data, and this separation time is the symbol period ($T_s$). The other parameter that is key to our discussion is the delay spread ($T_m$), which is a measure, typically a spacially averaged measure, of the time spread of the MPCs. The $T_m$ is dependent on the propagation environment, whereas $T_s$ is dependent on the wireless hardware. So, every wireless communication system can be regarded as something that occupies a unique point on the $T_mT_s$--plane.\\
\indent Wireless RXs use different models to approximate the mixture in the RHS of \eqref{eq11}. For better performance, the $r'(t)$ should be modelled as accurately as possible for different regions, or equally for the different classes of channels\footnote{The wireless channels are often defined in terms of the bandwidth of the data signal $s(t)$. For instance, if the bandwidth of $s(t)$ is small, the channel is said to be narrowband. The corresponding channel is said to be broadband, if the bandwidth of $s(t)$ is large. This definition however is colloquial and also incomplete. As described in \cite[Secs. I-A,B]{BasJ23}, in order to define wireless channels fully, two parameters: $T_m$ and $T_s$, or variants of them, are needed.}, on the $T_mT_s$-plane in Fig. \ref{fig:fig0}. For facts, the model:
\begin{flalign}\label{eq1}
\text{\textbf{Narrowband:}            } \quad \quad \quad r(t) &\approx hs(t)+w(t),&&
\end{flalign}
%
%
which approximates $r'(t)$ as a scaled version of $s(t)$, is deemed accurate for the RED region only, or colloquially when $T_m \lll T_s$, thus the name ``\textit{narrowband model}'' for \eqref{eq1}.\\
\indent In this model, the effect of the entire propagation environment between the TX and the RX is reduced to a single factor $h$, known also as the channel coefficient or channel gain or fading factor. It is widely known that, in this case, $h$ is the sum of the complex gains (i.e. $\gamma_n=\alpha _{n}e^{-j\phi _{n}}$) of the underlying MPCs. The RX samples $r(t)$ at regular intervals, and using the knowledge of $h$, the information is detected subsequently. The quality of the information detection is typically measured in terms of bit-error-rate (BER), and often affected by many factors including the modulation scheme, the channel coding scheme, the accuracy of the channel estimation, accuracy of the RX synchronization and the detection scheme\cite{Gold05}. 
\section{Deep Fading}\label{sec:deep_fading}
Let's consider the fading factor $h$ in \eqref{eq1}. It is widely known that, the typical PDF of the random process that gives rise to $h$ in a NLoS (non-line-of-sight) propagation environment looks like the bell curve shown in Fig. 2 \cite{Reudink74} and \cite[Sec. 5.4]{Molisch11}. It is a zero mean Gaussian variate with $\sigma_O^2$ variance and unimodal. This unimodality simply means that, it has a single peak. This unimodal PDF can be mathematically expressed as: 
\begin{align}\label{eq3}
	f_{h^i}(x) &= \frac{1}{\sqrt{2\pi\sigma_O^2}} e^{-\frac{x^2}{2\sigma_O^2}},
\end{align}
where $h^i=\text{Re}\{h\}$. Note that $h$ is complex, and $h^i$ and $h^q=\text{Im}\{h\}$ are independent and identically distributed. The PDF in \eqref{eq3} is such that as $\sigma_O^2$, otherwise the average strength of the fading factor, increases the PDF broadens and the peak reduces. As $\sigma_O^2$ decreases, the PDF thins, and the peak increases. The one and only degree of freedom available in this model is this $\sigma_O^2$.\\
\indent It is the Gaussianity in \eqref{eq3} that gives rise to the Rayleigh fading model widely used in wireless communication \cite{Reudink74}. This Rayliegh fading model is one of the pillars in modern wireless communication literature, and has been instrumental in analysing and predicting the performance of numerous technologies for many decades \cite{Foschini96,ZhTse03,YaMa13,Te99,Tarokh99,LoHe05,GoldCh97,Alouini99,Ali08}.\\
\indent In \cite{Foschini96}, Foschini uses the Rayleigh fading model to predict, in his own words, ``\textit{enormous}'' capacity gains of multi-elements antennas in fading. In \cite{ZhTse03}, Zheng et. al. exploits the simplicity of the Rayleigh fading model to formulate an elegant characterization of the optimal performance tradeoff of multiple antenna schemes. In \cite{Tarokh99}, Tarokh et. al. use Rayleigh fading model to derive a design criteria for space-time codes, and in \cite{LoHe05}, Love et. al. use the same model to design a limited feedback unitary precoding scheme for multiple-input-multiple-output (MIMO) spatial multiplexing systems. In \cite{GoldCh97,Alouini99}, Goldsmith et. al. use the Rayleigh model to design adaptive variable rate and variable power digital modulation schemes for wireless communication in fading. In \cite{Ali08}, Maddah-Ali et. al, use the Rayleigh model for fading to predict the multiplexing gain achievable in what they call MIMO X channels using interference alignment.\\
\indent Amidst all these, the bell curve in Fig. \ref{fig:fig1} signifies something that is profoundly unfavourable for wireless communication too: deep fading caused by the multipath. The deep fading is the most problematic phenomenon in wireless communication, which exerts its influence, often unfavourably, on every element of wireless communication systems. The peak at zero in Fig. \ref{fig:fig1} signifies the fact that the probability of $h$ being very small or being in deep fade is very high, which is highly unfavourable for wireless communication. On the other hand, tails signify the fact that $h$ can be large too, but it is highly unlikely.\\
\begin{figure*}[t]
	\centerline{\includegraphics*[scale=0.86]{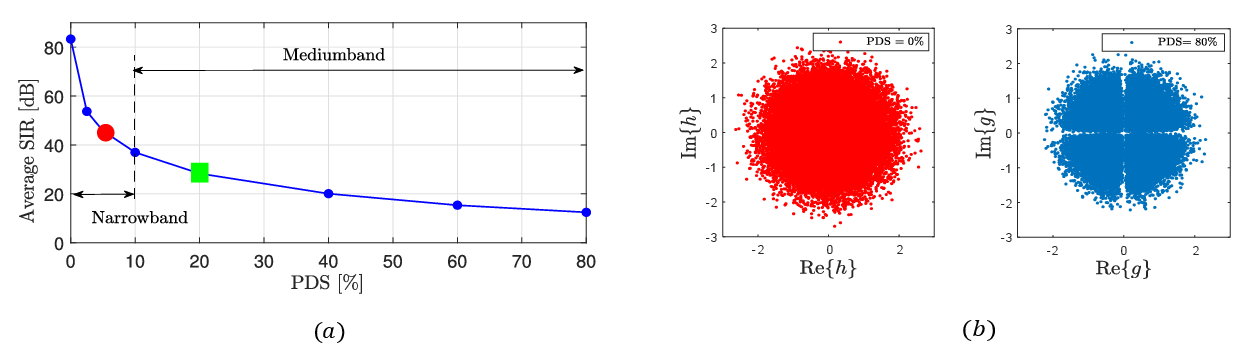}}
	\caption{a) Typical trend of average \text{SIR} with PDS. When PDS is less than 10\%, the narrowband model (i.e. \eqref{eq1}) is deemed to be satisfied, and the larger region, where PDS is greater than 10\%, is the mediumband region. The RED circle (i.e. $\text{PDS}=5\%$ ) represents a narrowband system, while the GREEN square (i.e. $\text{PDS}=20\%$ ) represents a mediumband system \cite{BasJ23}. b) Scatter plots showing the real and imaginary parts of $h$ in \eqref{eq1} and $g$ in \eqref{eq2} in NLoS propagation, where $\beta=0.22$ and $N=10$ (see Table \ref{table:1} for variable definitions). The low correlation between the real and the imaginary dimensions and, due to the ``\textit{\textbf{four leaf clover}}'' shape on the right, the effect of deep fading avoidance in $g$ are clearly visible.}\label{fig:fig11}
	\vspace{-5mm}
\end{figure*}
\indent As many researchers have correctly pointed out previously, it is not the lack of knowledge of $h$ at the RX, but the deep fading in $h$, that mainly adversely affects the detection performance, thus the overall quality, of wireless communication. Tse and Viswanath in \cite[pg. 70]{Tse04} elegantly articulate this important point saying, ``... \textit{the main reason why detection in fading channel has poor performance
is not because of the lack of knowledge of the channel }(i.e., $h$ in \eqref{eq1}) \textit{at the receiver. It is due to the fact that the channel gain is random and there is a significant probability that the channel is in a deep fade}''. The emerging mediumband wireless communication, uncovering some enlightening insights on the inner working of RF wireless channels, offers us new avenues to tackle this considerably problematic deep fading \cite{BasJ23,Bas22}. But, how does it do that? 
\section{Mediumband Wireless Communication}\label{sec:mediumband}
Due to historical reasons or for convenience, typically the model in \eqref{eq1} is attributed to the entire region on the $T_mT_s$-plane, where $T_m < T_s$ \cite{Sklar97}. However, \eqref{eq1} is a highly limiting case, and strictly speaking, cannot be found in reality. It is however approximately deemed accurate only in the small narrowband region marked in RED on the $T_mT_s$-plane in Fig. \ref{fig:fig0}\footnote{Also, by applying IFFT (Inverse Fast Fourier Transform) and FFT (Fast Fourier Transform) at the TX and the RX respectively, a set of similar equations can be obtained for the broadband region, which is in BLUE in Fig. \ref{fig:fig0}, as well \cite[Sec. 12.4]{Gold05}.}. The research has shown that, the model in \eqref{eq1} is wholly inadequate and inaccurate to use \eqref{eq1} to characterize the vast mediumband region in GREEN in between the narrowband and the broadband regions on the $T_mT_s$-plane \cite{BasJ23}.\\
\indent When the operating point moves into the mediumband region, primarily happening two things, the model in \eqref{eq1} morphs. Firstly, the strength, measured in terms of the average power, of the desired signal component reduces. Also, the energy that leaves the desired signal gives rise to an interference signal, which is a form of ISI, in additive form as \cite[eq. 13]{BasJ23}:
\begin{align}\label{eq2}
r(t) &\approx \underbrace{gs(t)+i(t)}_{r'(t)}+w(t),
\end{align}
where  $gs(t)$ is the new desired signal component, $g$ is the new desired fading factor, and $i(t)$ is the residual interference signal. Since it is the same total received power in $r'(t)$ that is now split into two signals, the quality of the channel in \eqref{eq2} measured in terms of average SINR or SIR is invariably lower than that of the channel in \eqref{eq1}, which is also evident in Fig. \ref{fig:fig11}-(a). In Fig. \ref{fig:fig11}-(a), the average SIR is the ratio of the average power of $gs(t)$ to the average power of $i(t)$ in decibels (dB).\\
\indent In Fig. \ref{fig:fig11}-(a), the RED circle (i.e. $\text{PDS}=5\%$) represents a narrowband system, while the GREEN square (i.e. $\text{PDS}=20\%$) represents a mediumband system. Here percentage Ddlay spread (PDS) denotes to the ratio of $T_m$ to $T_s$ as a percentage:
\begin{align}
	\text{PDS} &=\left(\frac{T_m}{T_s} \right) \times 100\%,
\end{align}
which is a simple metric that captures, as we call it, the degree of mediumband-ness of mediumband channels \cite{BasJ23}.\\ 
\indent One may interpret the two systems at RED circle and GREEN square in Fig. \ref{fig:fig11}-(a) in two different ways. On the one hand, for a fixed symbol period, the GREEN square represents a communication system operating in a relatively unfavourable propagation environment\footnote{Typically in digital wireless communication, a propagation environment is said to be favourable if the delay spread is low in comparison with the symbol period and the majority of receive power is concentrated in a few MPCs. Remember, it is not the absolute value of the delay spread, but the delay spread in comparison with the symbol period, that matters here.\label{xxx}} with a delay spread four times as wide as the delay spread of the system represented by the RED circle. Also, for a given propagation environment, which also means for a given delay spread, the mediumband system in GREEN square represents a system with a symbol rate four times as high as the symbol rate of the system represented by the RED circle. Since the symbol period can be made to be smaller than that of the RED system, the signalling can be done at a higher rate in mediumband systems enabling significantly higher data rate than that of narrowband systems. On the flip side, as the signalling rate increases or as the degree of mediumband-ness increases or as PDS increases, as shown in Fig. \ref{fig:fig11}-(a), the quality of the channel measured in terms of SIR decreases.\\    
\indent At first glance, the fact that SIR reduces steadily as PDS increases is bad news for the GREEN region, or in general for mediumband wireless communication. However, the story of mediumband does not end there. On the contrary to the widely held belief, it is not the SIR, or equally SINR, that dictates the performance of wireless communication in the mediumband, but the net effect of the SIR and another favourable effect known as the ``\textit{\textbf{effect of deep fading avoidance}}'' that could successfully counter the adverse effect of the increased level of ISI in \eqref{eq2}. This deep fading avoidance effect is such that it can significantly mitigate the unfavourable effects of ISI in all practically relevant signal-to-noise ratio (SNR) regions, as seen in Figs. \ref{fig:fig8} and \ref{fig:fig9}. Therefore, communicating in the mediumband is not a sacrifice of reliability for higher data rates, but if designed properly, can be made to be a ``\textit{\textbf{win-win}}'' situation \cite{BasJ23}. What is this effect of deep fading avoidance, and why does it matter?
\begin{figure*}[t]
	\centerline{\includegraphics*[scale=0.55]{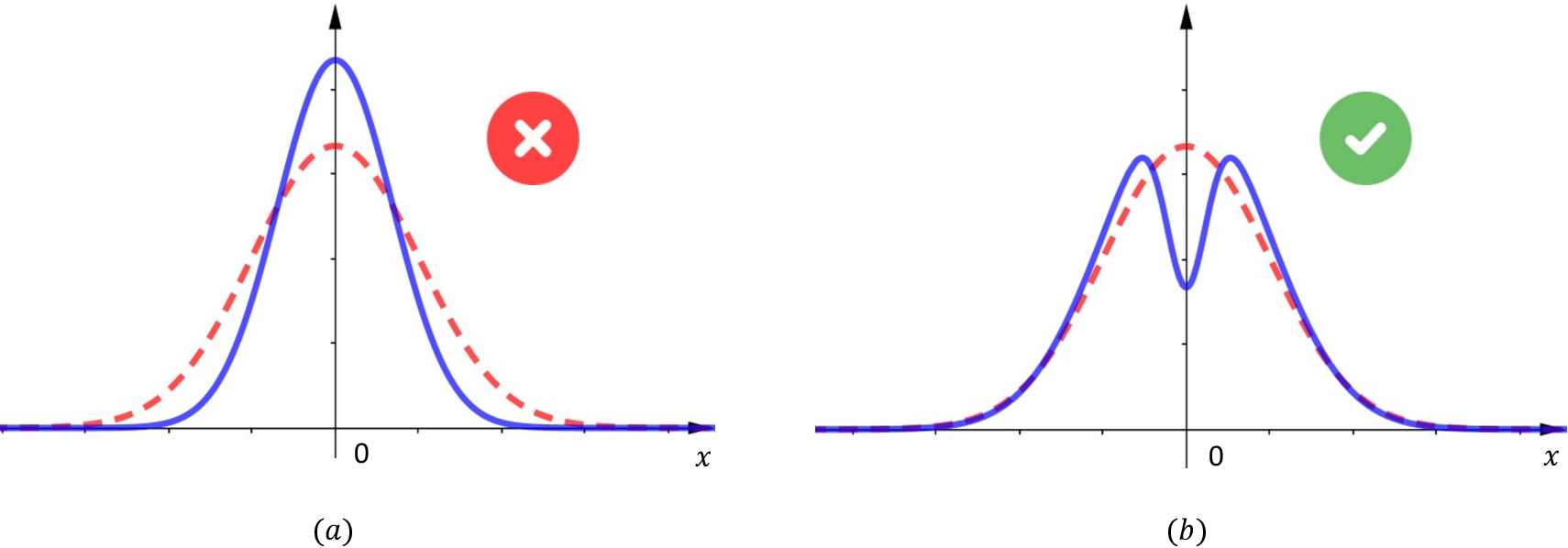}}
	\caption{On the right is a typical PDF of $g$ (only the real part) in a typical NLoS propagation environment \cite{BasJ23,Bas22}. Importantly, it is the same NLoS environment, which gives rise to the Gaussian PDF (in RED dotted line) for $h$ for narrowband systems, that gives rise to the bimodal PDF (in BLUE solid line) for $g$ for mediumband systems. It is the different operating points on the $T_mT_s$–plane that create this difference, but not different propagation environments.}\label{fig:fig2}
	\vspace{-2mm}
\end{figure*}
\section{The effect of Deep Fading Avoidance}\label{sec:dfa}
\begin{figure*}[t]
	\centerline{\includegraphics*[scale=0.83]{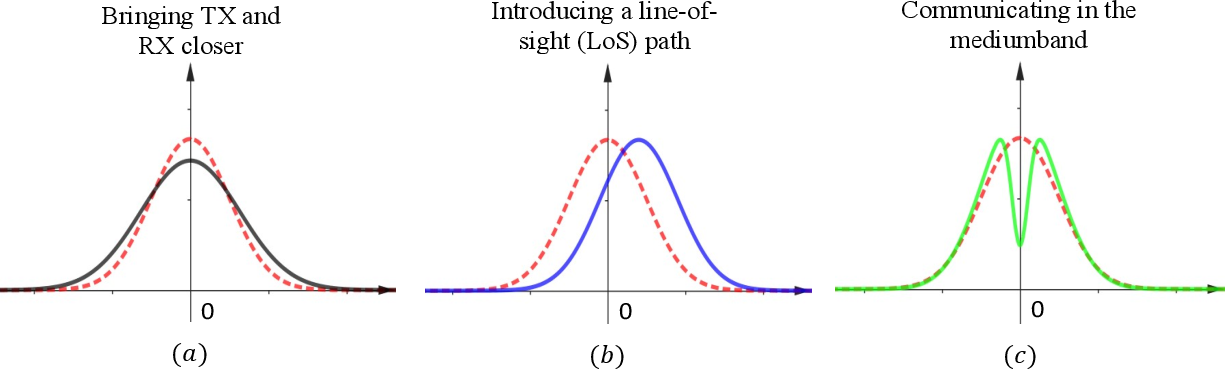}}
	\caption{Depictions of the resulting PDFs (only a single dimension) of the desired fading factor of several methods known to reduce deep fading.}\label{fig:fig31}
	\vspace{0mm}
\end{figure*}

It has turned out that, the desired fading factor $g$ in \eqref{eq2} is a complicatedly weighted sum of the complex gains of the underlying MPCs while $h$ is a regular sum \cite[eq. 15]{BasJ23}. As a result, the statistics of $g$ gives rise to refreshingly new perspectives. As the comparison of scatter plots of $g$ and $h$ in Fig. \ref{fig:fig11}-(b) clearly shows, the statistics of $g$ exhibit unique properties. In Fig. \ref{fig:fig11}-(b), one can make two immediate observations. Firstly, the radius of the scatter plot of $g$ (or the cloud) on the right is lower than that of the cloud on the left, which gives a hint that the average strength of $g$ is lower than that of $h$. Secondly, in contrast to the scatter plot of $h$, the cloud of $g$ is not circular, but has a unique ``\textit{\textbf{four leaf clover}}'' shape.\\
\indent When the average strength of the desired fading factor reduces, one may expect the PDF of $g$ to be, while maintaining the unimodality as shown in Fig. \ref{fig:fig2}-(a), just a thinner and taller version of the PDF of $h$. However, that is not what happens in the mediumband region. If the RX synchronization is achieved as described in \cite{BasJ23}, a typical PDF of $g$ in a NLoS propagation environment would look like the one shown in Fig. \ref{fig:fig2}-(b) \cite{BasJ23,BasSmith23}. Importantly, it is not a new propagation environment that gives rise to the PDF in Fig. \ref{fig:fig2}-(b) for $g$, but the same NLoS environment that would have given rise to the bell curve in Fig. \ref{fig:fig1} for $h$. It is the change of the operating point on the $T_mT_s$-plane, but not the propagation environment, that creates this difference. A process seems to have carved out a part of the PDF of $g$ from the top giving rise to a bimodal PDF. The depth and the width of the hole increase as the degree of mediumband-ness measured in terms of PDS increases. Also, it is this bimodality in the PDF of $g$ that gives rise to the ``\textit{four leaf clover}'' shape for the scatter plot of $g$ in Fig. \ref{fig:fig11}-(b). What does the hole at the top of the PDF of $g$ in Fig. \ref{fig:fig2}-(b) mean?\\
\indent On the one hand, this hole behaviour that is only prominent in the GREEN region but not in the RED region highlights the existence of a distinct region with unique properties on the $T_mT_s$-plane between the narrowband and broadband regions. So, the conventional teaching of wireless communication may need revisions to include the GREEN region along with its unique properties. On the other hand, the hole in the PDF of $g$ also signifies the fact that the probability of the desired fading factor in the mediumband channels, being in deep fade is now reduced.\\
\indent Among many methods, the way that mediumband wireless communication reduces deep fading is unique. Fig. \ref{fig:fig31} graphically depicts the resulting PDFs of several methods known to reduce deep fading. Bringing TX and RX closer is a simple, but often impractical, method to reduce deep fading. When bringing the TX and RX closer, it reduces the path loss increasing the average strength of the fading factor, otherwise $\sigma_O^2$. As shown in Fig. \ref{fig:fig31}-(a) in solid BLACK curve, this has a flattening effect on the peak of the PDF of the desired fading factor reducing the probability content around zero. Also, introducing a line-of-sight (LoS) path can reduce deep fading too. When a stable LoS path comes into presence, zero mean property of the PDF of $h$ is compromised moving the peak of the PDF away from zero, which in turn reduces the probability of deep fading as shown in Fig. \ref{fig:fig31}-(b) in solid BLUE curve. In both methods, the unimodality remains unchanged, whereas as shown in Fig. \ref{fig:fig31}-(c) in solid GREEN curve, when communicating in the mediumband, it is the bimodality of the PDF of $g$ that reduces deep fading.\\
\indent What is the extent of the effect of deep fading avoidance in mediumband channels for wireless communication?\footnote{Note that the effect of deep fading avoidance does not cease to exist immediately when the operating point moves into the broadband region. Depending on the nature of the propagation, one or more fading factors in the broadband channels could also be made to exhibit deep fading avoidance. A detail treatment on this can be found in \cite[Sec. VII]{BasJ23}}. 
\begin{table}[t]
	\begin{center}
		\caption{Parameters used for the simulations in Sec. \ref{sec:numerical_results}}
		\label{table:1}
		\begin{tabular}{|c|c|} 
			\hline
			\textbf{Parameter} & \textbf{Value}  \\
			\hline
			Symbol period ($T_s$) & 1[s] \\
			\hline  
			Delay spread ($T_m$) & changes to achieve different PDS. \\ 
			\hline
			Number of MPCs ($N$) & 10 \\
			\hline
			MPC index ($n$) & $n=0,1,\dots,9$  \\
			\hline
			Pulse shaping filter & SRRC at TX and RX \\
			\hline 
			Roll-off factor ($\beta$) & 0.22 \\
			\hline
			Span of pulse shaping filter & $12T_s$\\
			\hline
			$n$th Path delay & $U[0,T_m]$ \\
			\hline
			$n$th Path Phase & $U[0,2\pi]$ \\
			\hline
			$n$th average path power & $\propto 1/N$ \\
			\hline
			Modulation scheme & BPSK \\
			\hline
			Frame length & 100 bits\\ 
			\hline
		\end{tabular}
	\end{center}
	\vspace{-2mm}
\end{table}      
\section{Performance Analysis}\label{sec:numerical_results}
In order to get a sneak peak at the potential gain on offer, we herein report some emblematic results of a link-level simulation study conducted on MATLAB. A generic NLoS wireless propagation scenario in a rich scattering environment for a single-input-single-output (SISO) wireless link with BPSK modulation is considered. As in \cite{3gpp25996} for urban microcells, the propagation delays corresponding to $N$ MPCs are drawn from a uniform distribution: $U[0,T_m]$, where $T_m$ is the delay spread. The path phases are drawn from a uniform distribution: $\phi_n \sim U[0,2\pi]$. Corresponding to a severe multipath scenario, the average path powers are assumed to be equal, and is normalized such that all path powers are summed to unity. Note that, it is the average path powers that are equal, but the $n$th instantaneous path power is drawn from a Rayleigh distribution with an appropriate scale parameter \cite{Molisch14}. These assumptions and normalizations mean that if a wireless system is operated in the narrowband regime under these assumptions and normalizations, the corresponding fading factor $h$ is such that $\mathcal{E}\{|h|^2\}=1$, thus $\mathcal{E}\{(h^i)^2\}=\mathcal{E}\{(h^q)^2\}=0.5$. By changing $T_m$  appropriately while keeping $T_s$ fixed, mediumband channels with different PDS are obtained, and other key parameters are listed in Table. \ref{table:1}.\\
\begin{figure}[t]
	\centerline{\includegraphics[scale=0.65]{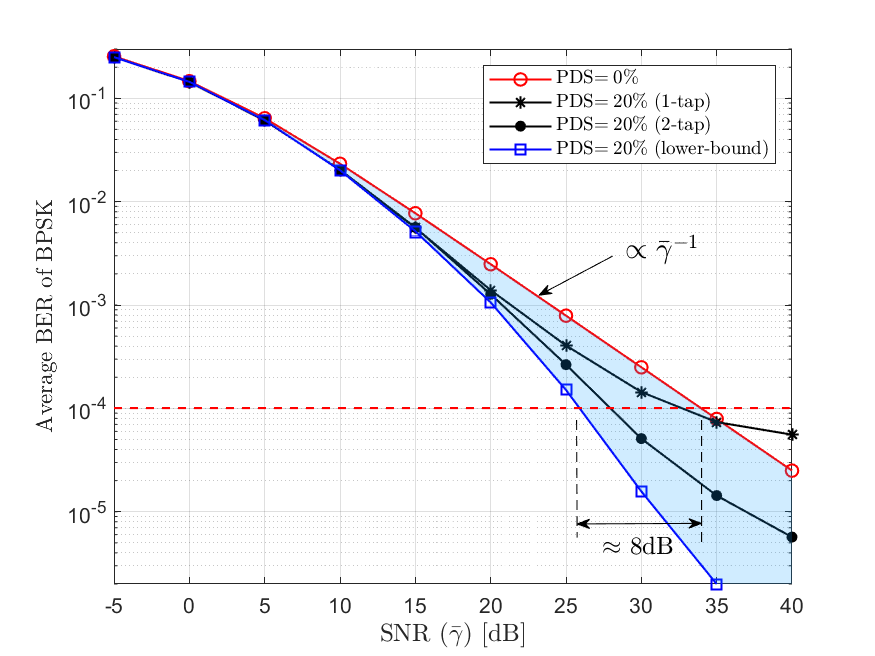}}
	\caption{Average BER of BPSK modulation against the received SNR $\bar{\gamma}$ for PDS=$20\%$ in NLoS propagation. The lower bound is from \cite[eq. 16]{BasJia23} with values for $K$, $\sigma_I^2$ and $\sigma_O^2$ from Table. \ref{table:2}.}\label{fig:fig8}
	\vspace{-6mm}
\end{figure}
\indent Figs. \ref{fig:fig8} and \ref{fig:fig9} show the average uncoded BER of BPSK modulation for two mediumband channels with PDS=$20\%$ and PDS=$60\%$ respectively. Here for instance, for a given symbol period, PDS=$20\%$ means that the $T_m$ is a fifth of the symbol period, and PDS=$60\%$ means that the delay spread is three fifth of the symbol period. For comparison, BER performance of the ideal narrowband channel with no ISI in \eqref{eq1} is shown in both figures (see the RED curve). It is a well known performance limit and the diversity order achieved is one \cite{Tse04}. This ideal ISI free circumstance is a limiting case, and can only be achievable when $T_m \rightarrow 0$, thus the label PDS=$0\%$. Consequently, PDS=$20\%$ means a moderately unfavourable propagation environment with a delay spread 8 times as wider as the delay spread of the case of PDS=$0\%$\footnote{Here it is assumed that, for practical purposes, $T_m$ being 1/40th of $T_s$ is sufficient to approximately achieve the ISI free scenario in RHS of \eqref{eq1}}. Also, PDS=$60\%$ means a more unfavourable propagation environment with a delay spread 24 times as wider as the delay spread of the case of PDS=$0\%$. Yet, as shown in Figs. \ref{fig:fig8} and \ref{fig:fig9}, due mainly to the deep fading avoidance, BER performance of PDS=$20\%$ and PDS=$60\%$ significantly outperform the BER performance of ISI free ideal narrowband channel described in \eqref{eq1}. In all simulations herein, the average receive SNR denoted by $\bar{\gamma}$ is the ratio of the average power of $r'(t)$ to the average power of the noise signal, $w(t)$, where $r'(t)$ is defined in \eqref{eq11}.\\       
\indent It is still an open problem as to what the optimum detection algorithm for channels in the mediumband region is, but Figs. \ref{fig:fig8} and \ref{fig:fig9} report BER performance, against the receive SNR, $\bar{\gamma}$, of two suboptimal detection schemes alongside the BER of the ideal narrowband channel described in \eqref{eq1} (RED curve) and a lower bound (BLUE curve). In the first suboptimal scheme, multipath signals are modelled into a single tap (thus the name ``1-tap'') as described in \cite[Theorem 1]{BasJ23} with symbol-by-symbol detection, and in the second scheme, the multipath signals are modelled into two taps (thus the name ``2-tap'') as described in \cite[Corollary 1]{BasJ23} with successive interference cancellation (SIC) \cite{BasJia23}. As detailed in \cite{BasJia23}, the lower bound is obtained from a hypothetical channel shown in \eqref{eq2}, but without the additive ISI term, $i(t)$.\\
\begin{figure}[t]
	\centerline{\includegraphics[scale=0.65]{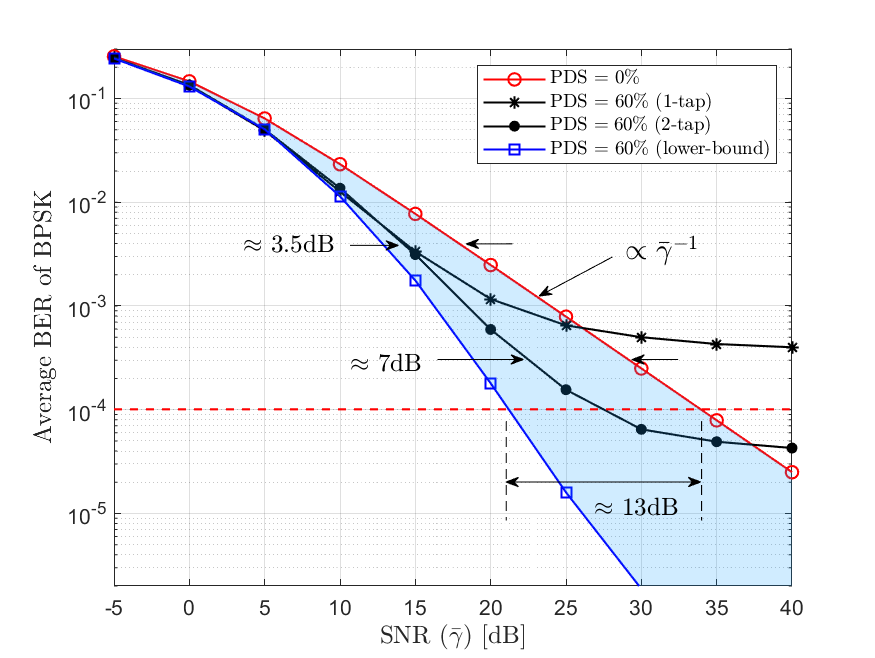}}
	\caption{Average BER of BPSK modulation against the received SNR $\bar{\gamma}$ for PDS=$60\%$ in NLoS propagation. The lower bound is from \cite[eq. 16]{BasJia23} with values for $K$, $\sigma_I^2$ and $\sigma_O^2$ from Table. \ref{table:2}.}\label{fig:fig9}
	\vspace{-6mm}
\end{figure}
%
%
\indent In Fig. \ref{fig:fig8} for PDS=$20\%$, BER of the 1st scheme (i.e., ``1-tap'') outperforms the ideal ISI free channel up to SNR=35dB in some cases by 2.5dB. After that, BER seems to saturate. It is because, the mediumband channel is a battleground of two opposing
phenomena: ISI and the deep fading avoidance. At low SNR, it is the deep fading avoidance that is dominant successfully countering the adverse effect of ISI and garnering several dBs of BER gain up to SNR=35dB. However, as SNR increases the effect of ISI eventually kicks in and dominate the effect of deep fading avoidance leading to a saturation of BER after SNR=35dB. Furthermore, the BER gain of the 2nd scheme (i.e., ``2-tap'') over the BER of the channel in \eqref{eq1} is, in some cases, more than 5dB. The BER lower bound (BLUE curve) drops at a faster rate, due to the effect of deep fading avoidance in the corresponding fading factor $g$, than the BER of the ideal ISI free channel in \eqref{eq1}. The BER achieved by ``2-tap'' scheme closely follows the lower bound lending credence to the claim that this lower bound is tight.\\
\indent The error floor in both ``1-tap'' and ``2-tap'' schemes in Fig. \ref{fig:fig9} is relatively high. Because, PDS=$60\%$ represents a significantly unfavourable channel for wireless communication, and the residual interference (or remaining ISI) that cannot be accounted for in these detection schemes is high. However, BER is significantly better than that of the RED curve up to SNR of 37dB (in ``2-tap'' case), in some cases by 7dB, due mainly to the effect of deep fading avoidance. The BER would be even better if SIC is employed with ``3-taps''. Importantly, the lower bound in Fig. \ref{fig:fig9} drops at even a faster rate, but is slightly way off from the BER curves of ``1-tap'' and ``2-tap'' schemes. It is not yet clear if practical detection schemes can achieve this lower bound for such highly unfavourable propagation conditions, but certainly in general, the lower bounds in both Figs. \ref{fig:fig8} and \ref{fig:fig9} function as good benchmarks. For instance, from these lower bounds, we can hypothesis that communicating in mediumband channels of PDS=$20\%$ and PDS=$60\%$ can achieve uncoded BER of $10^{-4}$ with staggering 8dB and 13dB respectively less receive SNR than that in the ISI-free narrowband channel. What do the shaded areas in Figs. \ref{fig:fig8} and \ref{fig:fig9} mean?\\
\indent The shaded area bounded by the BER of the narrowband channel in \eqref{eq1} and the lower bound may be termed as the ``\textit{\textbf{playing area}}'' for detection algorithms for mediumband channels. As shown in Fig. \ref{fig:fig9} for PDS=$60\%$, typically the shaded area increases as PDS increases. It is because, as PDS increases, owing to the increased effect of deep fading avoidance in $g$, the lower bound drops at even a faster rate.\\
\indent As a rule of thumb, the larger the shaded area, the higher the effect of deep fading avoidance in $g$. That is to say, the more unfavourable a propagation environment for wireless communication, the higher the potential gain by communicating in the mediumband.
\section{A General and Unified Model}\label{sec:statistical_model}
\begin{figure*}[t]
	\centerline{\includegraphics*[scale=0.72]{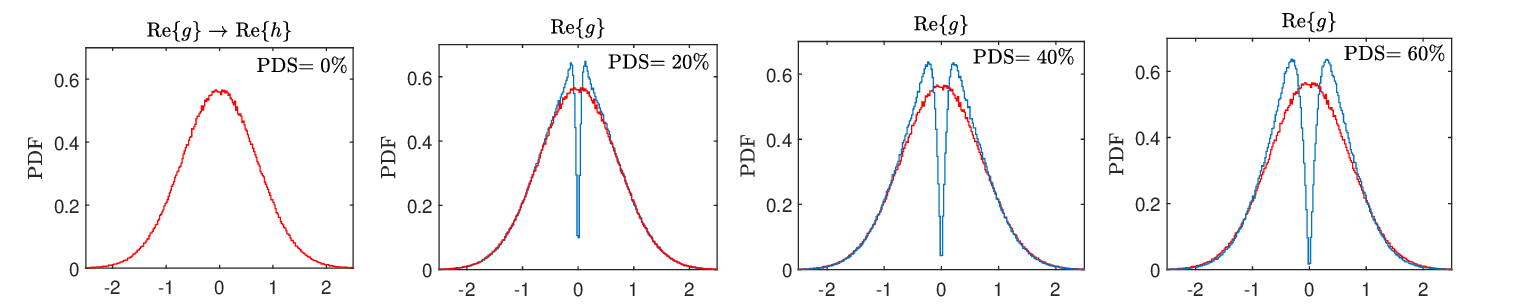}}
	\caption{The evolution of the hole behaviour in the marginal PDF of $g$ (i.e., of Re$\{g\}$ and Im$\{g\}$) as PDS increases. This is from 1 million samples each from a detailed link-level simulation using the parameters in Table \ref{table:1}.}\label{fig:fig41}
	\vspace{0mm}
\end{figure*}

\begin{figure*}[t]
	\centerline{\includegraphics*[scale=0.85]{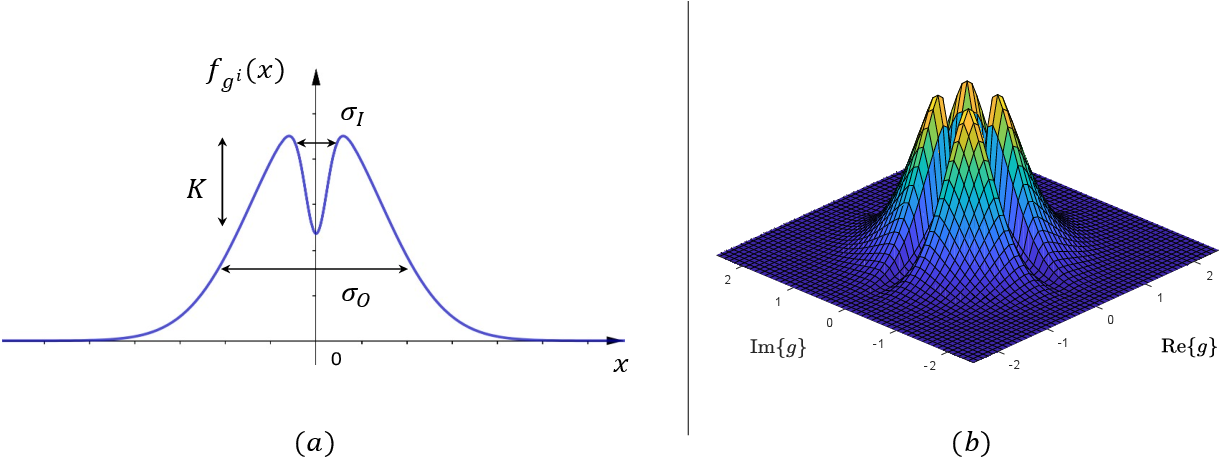}}
	\caption{a) A graphical representation of the model for the PDF of $g$ in (6a). Here $g^i$ denotes the real part of $g$. Note that $K, \sigma_I$ and $\sigma_O$ are parameters that capture the depth and the width of the hole and the outer width of the PDF respectively, but do not measure the distances exactly as shown here. b) A depiction of $f_{g}(x,y)=f_{g^i}(x)f_{g^q}(y)$, which is the complex PDF of the desired fading factor $g$ in mediumband wireless channels. The four peaks are due to the bi-modality in the marginal PDF of Re$\{g\}$ and Im$\{g\}$.}\label{fig:fig4}
	\vspace{0mm}
\end{figure*}

The extent of the effect of deep fading avoidance changes as PDS changes. As shown in Fig. \ref{fig:fig41}, as PDS increases, the depth and the width of the dip in the PDF of $g$ increase.  This statistical behaviour of $g$, particularly the bimodal nature, can be modelled very accurately in an analytically amenable form as a weighted difference of two Gaussian kernels \cite{BasSmith23}:
\begin{subequations}\label{eq6}
\begin{align}
	f_{g^i}(x) &= \frac{e^{-\frac{x^2}{2\lambda_0^2}}-Ke^{-\frac{x^2}{2\lambda_1^2}}}{\sqrt{2\pi\lambda_0^2} - K\sqrt{2\pi\lambda_1^2}},\\
	\nonumber \\
	\lambda_0 &= \sigma_O, \qquad \quad \lambda_1 = \sqrt{\frac{\sigma_O^2\sigma_I^2}{\sigma_O^2+\sigma_I^2}},
\end{align}
\end{subequations}
where $g^i=\text{Re}\{g\}$. Note that, $g^i$ and $g^q=\text{Im}\{g\}$ are identically distributed, and the research so far also gives credence to the fact that they are independent too. As shown in Fig. \ref{fig:fig4}, in contrast to the bell curve in Fig. \ref{fig:fig1}, three primary parameters are needed to accurately model the hole (or dip) in the PDF. The parameters $K$ and $\sigma_I$ respectively capture the depth and width of the hole, where $\sigma_O,\sigma_I$, $K$ are all non-negative, and $\sigma_O \gg \sigma_I$. One can clearly see that as $\sigma_I \rightarrow 0$,  the prominence of the hole gradually diminishes, reducing  $f_{g^i}(x)$ to the Gaussian PDF in \eqref{eq3}. So, the model in \eqref{eq6} is a general case, which includes the classical Gaussian variates that gives rise to the widely used Rayleigh fading model as a special case. As described in \cite{BasSmith23}, this Gaussian-hole variates, or $g^i$ (and equally $g^q$), can be generated easily enabling rapid simulation of mediumband channels. Furthermore, this model seems to be able to predict the link-level performance of mediumband wireless communication quite successfully\cite{BasJia23}.\\
\indent Table \ref{table:2}, furthermore, summarizes the estimated parameter values for $K$, $\sigma_I^2$ and $\sigma_O^2$ for several PDSs\footnote{A similar set of estimated values for these parameters has also been reported in \cite{BasSmith23,BasJia23}, but the values reported herein are more accurate.}. Note that, these values are evaluated for a normalized channel meaning that the average path powers are normalized to one. As we can see both $K$ and $\sigma_I^2$ increase as PDS increases, which means that as PDS increases the depth and the width of the dip in the PDF of $g$ increase. Typically the deeper and the wider the dip, the higher the effect of the deep fading avoidance in mediumband channels.       
\vspace{0mm}
\section{Challenges and Wider Impact}\label{sec:impact_problems}
In wireless communication, it is the ``\textit{statistics of the fading factors}'', but not the ``\textit{knowledge of the fading factors}'' that is more important. As the evidence presented in this paper shows, communicating in the mediumband is a new vantage point to alter the statistics of the fading factors arisen in wireless communication. Communicating optimally in the meadiumband is however not without challenges. As for a basic SISO mediumdband wireless communication link, which is the scenario of a single TX and a RX, a few areas where further research is needed can be summarised as:
\begin{table}[t]
	\begin{center}
		\caption{Typical values for the three main parameters in the PDF of $\text{Re}\{g\}$ for Different PDSs.}
		\label{table:2}
		\begin{tabular}{|c|c|c|c|} 
			\hline
			PDS $[\%]$ & $K$  & $\sigma_I^2$ & $\sigma_O^2$\\
			\hline
			0 & 0 & 0 & 0.5000\\
			\hline 
			20 & 0.9218 & 0.0008 & 0.4818\\
			\hline
			40 & 0.9336 & 0.0031 & 0.4580\\
			\hline  
			60 & 0.9502 & 0.0074 & 0.4338\\ 
			\hline
			80 & 0.9668 & 0.0131 & 0.4054\\ 
			\hline
		\end{tabular}
	\end{center}
	\vspace{-4mm}
\end{table}
\begin{itemize}
\item For all the results in \cite{Bas22,BasJ23,BasSmith23,BasJia23} and in this paper, the perfect carrier phase and symbol timing synchronizations are assumed. Typically, mediumband wireless communication requires very accurate RX synchronization, specially symbol timing, for optimal performance. If the RX synchronization in terms of carrier phase and symbol timing are suboptimal, the performance should reduce gracefully. The extent of the performance degradation of suboptimal RX synchronization (specially symbol timing) should be studied.
\item As described in \cite[Fig. 7]{BasJ23}, harnessing deep fading avoidance for one-dimensional modulations (such as BPSK) is straightforward. The effect of deep fading avoidance is lightly dependent on the modulation through the roll-off factor (i.e., $\beta$) only, but not on the modulation type. However, two-dimensional modulations (such as QPSK) could impose constraints on symbol timing synchronization, which could affect the achievable deep fading avoidance unfavourably. So, understanding the full extent of the achievable deep fading avoidance for two-dimensional modulations requires further research.   
\item As reported in Figs. \ref{fig:fig8} and \ref{fig:fig9}, the effect of deep fading avoidance gives rise to significant performance gains over the ISI free Rayleigh fading channel, and have the potential to achieve even more gains as predicted by lower bounds. In some cases, the predicted gain (i.e., the lower bound) can be approachable (see Fig. \ref{fig:fig8}), but in some cases (see Fig. \ref{fig:fig9}), is way off. Further research on detection schemes (e.g. machine learning based) with reasonable channel state information (CSI) requirement is needed.
\item It is the mediumband channels where the effect of deep fading avoidance is most prominent and most influential, but the deep fading avoidance effect does not cease to exist immediately as $T_m$ goes beyond $T_s$, that is the broadband region. Depending on the nature of the propagation, specially the delay distribution of dominant MPCs, one or more coefficients in the discrete-time version of the broadband channels could exhibit deep fading avoidance \cite[Sec. VII]{BasJ23}. As a result, fading factors in OFDM or SC-FDM subchannels seem to experience deep fading avoidance too, but further research is needed to understand the full extent of the effect of deep fading avoidance in broadband channels.
\end{itemize}         
\indent The impact of communicating in the mediumband for wider wireless communication may be deep and wide. When the constituent links in systems like MIMO, cognitive radio, relaying, reflective intelligent surfaces, space-time coding, limited feedback, non-orthogonal multiple access, integrated sensing and communication, and countless emerging application areas of RF wireless communication are made to operate in the mediumband, these systems are expected to offer significant performance gains and create new opportunities.\\ 
\vspace{-5mm}
\balance
\section{Conclusion}\label{sec:conclusion}
Recent research has shown sufficient and conclusive evidence to support that RF wireless channels, specially that fall in the transitional region between the narrowband and broadband on the $T_mT_s$-plane, have many unique  properties including their inherent ability to avoid deep fading. The enlightening insights drawn into the inner working of the RF wireless channels include, among other things, the ability of the deep fading avoidance to counter the adverse effect of ISI garnering significant link level performance. The research has also been very productive in quantifying and modelling this effect giving us powerful statistical models. The potential impact of this new class of channels and associated models may be deep and wide, in particular as an enabling communication technology. Further research is needed to explore, both qualitatively and quantitatively, how the communicating in the mediumband affects the physical layer (PHY) of countless areas of RF wireless communication. The tools and the insights presented in this paper and its key references may be useful and a good starting point for those future endeavours. 
\vspace{0mm}
%

%
%
%
\begin{figure*}[t]
	\centerline{\includegraphics*[scale=0.85]{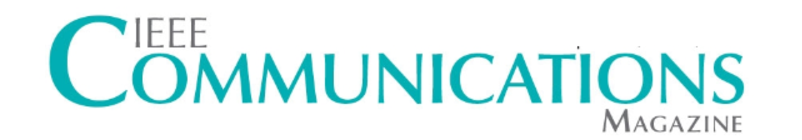}}
	\vspace{0mm}
	\hrule
\end{figure*}
\end{document}